\documentclass[aps,showpacs,twocolumn,aps,amsfonts,superscriptaddress,floatfix,nofootinbib]{revtex4}
\usepackage{amsmath}    
\usepackage{graphicx}   
\usepackage{verbatim}   
\usepackage{color}      
\usepackage{subfigure}  
\usepackage{hyperref}   
\usepackage{amsfonts}

\usepackage{subfigure}
\renewcommand\Re{\operatorname{Re}}
\renewcommand\Im{\operatorname{Im}}

\usepackage{amsmath}
\usepackage{amsfonts}
\newcommand{\G}{{\mathbb{G}}}

\newcommand{\GO}{{\mathbb{G}_0}}

\newcommand{\Td}{{\mathbb{T}_2}}
\newcommand{\To}{{\mathbb{T}_1}}
\newcommand{\rr}{\mathbf{r}}

\newcommand{\T}{\mathbb{T}}

\begin{document}

\title{Linear response relations in fluctuational electrodynamics}

\author{Vladyslav~A. Golyk}
\affiliation{Massachusetts Institute of Technology, Department of
  Physics, Cambridge, Massachusetts 02139, USA}
\author{Matthias Kr\"uger}
\affiliation{Max-Planck-Instit\"ut f\"ur Intelligente Systeme, Heisenbergstr. 3, D-70569 Stuttgart, Germany, and
IV. Institut f\"ur Theoretische Physik, Universit\"at Stuttgart,
Pfaffenwaldring 57, D-70569 Stuttgart, Germany}
\author{Mehran Kardar}
\affiliation{Massachusetts Institute of Technology, Department of
 Physics, Cambridge, Massachusetts 02139, USA}

\begin{abstract}
Near field radiative heat transfer and vacuum friction are just two
instances of topics of technological and fundamental interest studied via
the formalism of fluctuational electrodynamics.
From the perspective of experiment and simulations, it is hard to precisely control 
and probe such non-equilibrium situations.
Fluctuations in equilibrium are easier to measure, and typically can be related
to non-equilibrium response functions by Green-Kubo relations. 
We consider a collection of arbitrary objects in vacuum, perturbed by changing
the temperature or velocity of one object.
Developing a method for computation of higher order correlation functions
in fluctuational electrodynamics, we explicitly compare linear response
and equilibrium fluctuations. 
We obtain a Green-Kubo relation for the radiative heat transfer, 
as well as a closed formula for the vacuum friction in
arbitrary geometries in the framework of scattering theory. 
We comment on the signature of the radiative heat conductivity in 
equilibrium fluctuations.
\end{abstract}


\maketitle
\section{Introduction}\label{section1}
The theoretical analysis, simulation or experimental measurement of out of equilibrium quantities is important and challenging. The
inability to employ fundamental concepts of equilibrium statistical physics,
such as free energy or entropy, can make theoretical analysis laborious. 
Experimentally, it can be tedious to maintain the system in a well controlled non-equilibrium state. Equilibrium quantities are often easier to access. 
For example, the mean square displacement of a Brownian particle in equilibrium 
is well amenable to measurements, whereas the direct evaluation of the Brownian particle's mobility is generally more difficult~\cite{Dhont}. The two
quantities are linked by the well known Einstein linear response relation. 
More generally, linear response relations are helpful in understanding and
quantifying non-equilibrium properties in terms of equilibrium fluctuations. 
The Green-Kubo (GK) relation~\cite{Green, Kubo} allows to obtain e.g.
thermal~\cite{GKthermal} and electric~\cite{GKelectric} transport coefficients 
or the sheer viscosity~\cite{GKshear}, by connecting the linear transport 
coefficients to time integrals of equilibrium correlation functions of the fluxes associated with conserved densities, and has found applications in the context of molecular dynamics simulations.
For example, the heat
conduction coefficient $\kappa$, can be expressed in terms of the heat flux~\cite{KuboBook} as
\begin{equation}\label{GK}
\kappa=\lim_{t'\rightarrow\infty}\lim_{V\rightarrow\infty} \frac{1}{k_B T^2V}\int_0^{t'} dt \langle \mathcal{J}(t)\mathcal{J}(0)\rangle^{eq},
\end{equation}
where $V$ and $T$ are the volume and  temperature of the system, respectively, and $\mathcal{J}(t)$ is the total heat flux in the direction of the temperature gradient. We denote averages in equilibrium by $\langle\dots\rangle^{eq}$, while non-equilibrium averages are indicated as $\langle\dots\rangle$. Note that $\langle \mathcal{J}(t)\rangle^{eq}=0$.

Another example that has proven useful in simulation-analysis~\cite{SimKirkwood}  is the so-called Kirkwood formula, expressing the friction $\gamma$ of a particle~\cite{Kirkwood, KuboFDT, Volokitin2}  (quoting the result in one dimension) as
\begin{equation}\label{Kirkwood}
\gamma=\frac{1}{k_B T }\int_{0}^\infty dt  \langle \delta F(t) \delta F(0)\rangle^{eq},
\end{equation}
where $\delta F(t)=F(t)-\langle F(t)\rangle^{eq}$ is the fluctuating part of the instantaneous total force $F(t)$ acting on the particle (this notation is used for all observables in the following). 

In this paper, we study linear response relations for the quantum thermal fluctuations of the electromagnetic field, which are related to
radiative heat transfer $H$~\cite{vanHove} and the Casimir force $\bf F$~\cite{Casimir}. By explicitly computing correlation functions of these quantities for a collection of arbitrary objects in vacuum, we identify them with previously found
non-equilibrium expressions for  radiative heat transfer and  non-equilibrium Casimir force in Sec.~\ref{section22}, thereby obtaining a GK matrix for  the
heat conductivities. The non-equilibrium force is in turn related to the equilibrium correlation of $\bf F$ and $H$. We also explicitly confirm the validity of the Kirkwood relation in Sec.~\ref{section41}, thereby providing a closed form expression for the vacuum friction for a collection of arbitrary objects. We finally give a relation for the change in heat absorption upon changes in velocities, explicitly confirming the Onsager theorem. We close with a discussion of experimental relevance and summary of our findings in Sec.~\ref{section5}.

While Eq.~\eqref{Kirkwood} is an example of the fluctuation dissipation 
theorem~\cite{Reynaud} (position and force are conjugate
variables in the Hamiltonian), Eq.~\eqref{GK} is obtained from taking the limit of small spacial variation of thermodynamic driving forces (e.g. temperatures gradients)~\cite{KuboBook}. The case of radiative heat transfer is hence different as we consider disjoint objects. Our methods allow for an explicit computation of the analog of Eq.~\eqref{GK}, which is a useful check of linear response and provides additional insight into radiative transfer.

\section{Perturbing temperature}\label{section22}
\subsection{Radiative heat transfer}\label{section2}
\begin{figure}\centering
\includegraphics[width=8.1 cm]{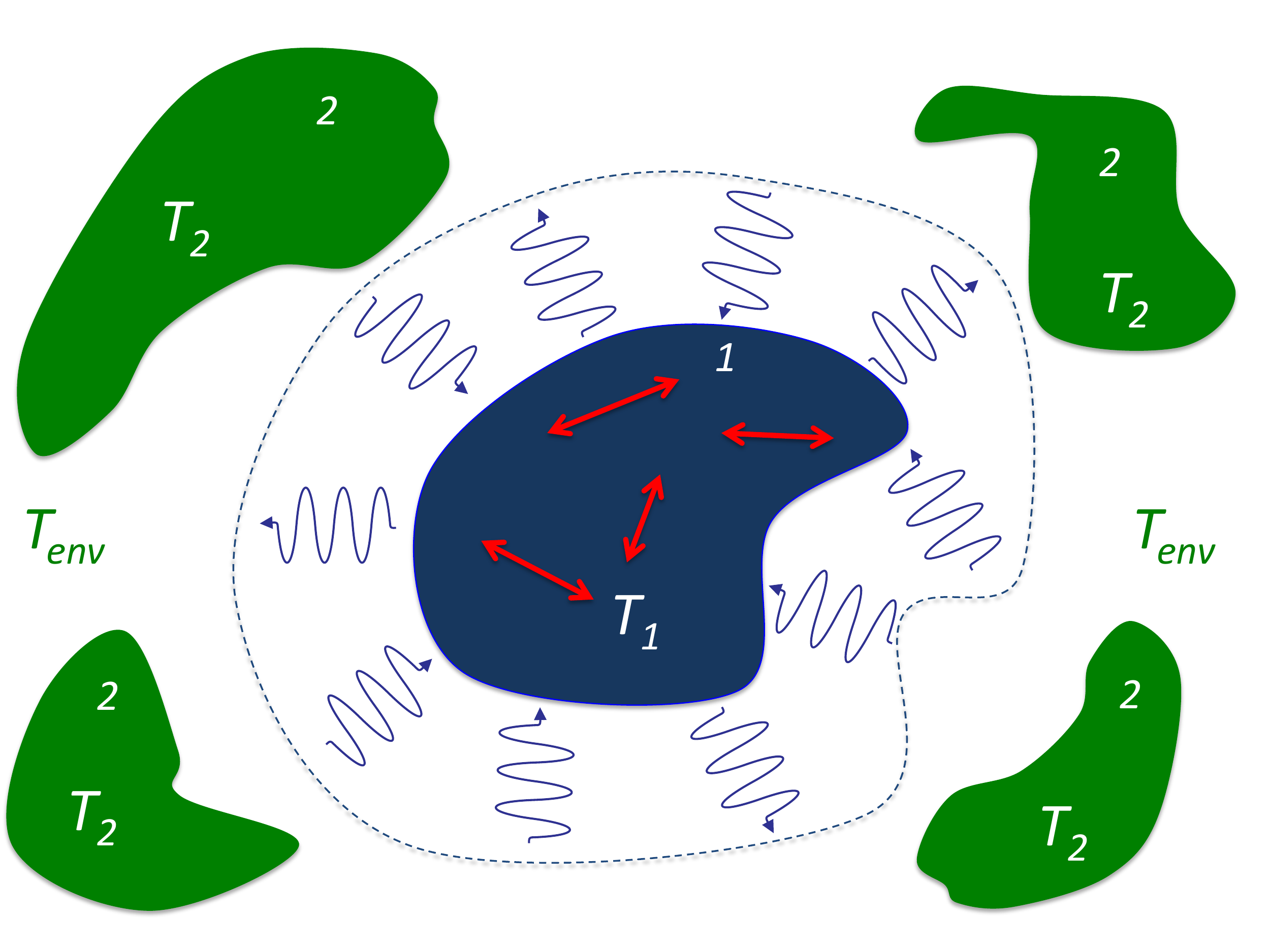}
\vspace*{-0.5cm}
\label{fig1} \caption{(color online)
The system under consideration consists of two (possibly multi-component) 
entities (blue and green). In equilibrium with $T_1=T_2=T_{env}$, the average heat absorbed by object $1$ (illustrated by the blue arrows)  is zero, and the net
force on it is the equilibrium Casimir force. 
If $T_1$ or $T_2$ slightly deviate from equilibrium, the finite heat absorption 
and the non-equilibrium Casimir force are given by Eqs.~\eqref{result1} and~\eqref{result2} respectively.
}\label{fig}
\end{figure}

Consider an arrangement of $N$ arbitrary objects such that $n$ of them are 
held at one set of conditions (temperature $T_1$ and velocity 
${\bf v}_1$), while the remaining $N-n$ objects are at slightly different conditions ($T_2$, ${\bf v}_2=0$). In the following we will denote the two groups by 
$\{\alpha,\beta\}=1,2$, keeping in mind that each entity can be made up of disconnected pieces. 
This collection is immersed in a vacuum at temperature $T_{env}$. 
Starting from the equilibrium situation with $T= \{T_\alpha\}=T_{env}$ and ${\bf v}_1={\bf v}_2=0$, we first introduce a small perturbation in the temperature of one of the objects (see Fig.~\ref{fig}), aiming to connect the corresponding  linear heat transfer coefficient to the fluctuations of the heat flux in equilibrium, in analogy to Eq.~\eqref{GK}. While the former has been derived in Ref.~\cite{LongPaper}, the latter will be found below. 

The total radiation energy $H^{(\beta)}$ absorbed by object(s) $\beta$ can be  written as an integral over the volume(s) $V_\beta$ of
the local work which is the product of the electric field $\bf E$ and current $\bf J$~\cite{Jackson} at point ${\bf r}$ and time $t$, leading to
\begin{equation}\label{CurrentFlux}
\begin{split}
H^{(\beta)}(t)=\int_{\rr\in V_\beta}d^3 \textbf{r} \{ E_i(\rr,t), J_i (\rr,t)\}_S\,.
\end{split}
\end{equation}
This expression can be recast as the surface integral of the Poynting vector through the Poynting theorem~\cite{LongPaper,Jackson}. 
We use the Einstein summation convention throughout, which implies summation over the vector index $i$ in Eq.~\eqref{CurrentFlux}, and
$\{A,B\}_S\equiv({A}{B}+{B}{A})/2$ is the symmetrized product of the generally non-commuting quantum operators. 
Note that $\langle H^{(\beta)}(t)\rangle^{eq}=0$.

The correlations between fluctuations of $H^{(\alpha)}(t)$ in equilibrium can be formally written (note that $\int_0^\infty dt
\langle \{{A}(t),{B}(0)\}_S\rangle^{eq}=\int_0^\infty dt\langle {A}(t){B}(0)\rangle^{eq}$, making  symmetrization needless on the left
hand side) as
\begin{equation}
\begin{split}\label{RHS}
&\int_0^\infty dt\langle H^{(\alpha)} (t) H^{(\beta)}(0) \rangle^{eq}=\int_0^\infty dt \iint_{\substack{\rr\in V_\alpha\\\rr'\in V_\beta }}d^3 \rr d^3 \rr' \\ &\langle \{E_i (\rr,t), J_i (\rr,t)\}_S\{E_j (\rr',0), J_j (\rr',0)\}_S \rangle^{eq}\,.
\end{split}
\end{equation}
The spacial integrals are restricted to the corresponding volumes according to Eq.~\eqref{CurrentFlux}. Equation~\eqref{RHS} contains a four-point correlation function of the electric field (noting the linear relation between $\bf E$ and $\bf J$ in Eq.~\eqref{current} below). 
Given the Gaussian distribution of the electric field,  Eq.~\eqref{RHS} can be rewritten in terms of time-ordered two-point correlation functions via Wick's theorem,
\begin{equation}
\label{eq:Wick}
\begin{split}
&\int_0^\infty dt\langle H^{(\alpha)}(t) H^{(\beta)}(0) \rangle^{eq}=\int_0^\infty dt\iint_{\substack{\rr\in V_\alpha\\\rr'\in V_\beta }}d^3 \rr d^3 \rr'
\\
&\times\left[\langle E_i (\rr,t)  E_j (\rr',0) \rangle^{eq}\langle J_i (\rr,t)  J_j (\rr',0) \rangle^{eq}\right. \\
&\left.+\langle E_i (\rr,t)  J_j (\rr',0)\rangle^{eq}\langle J_i (\rr,t) E_j (\rr',0) \rangle^{eq}\right]\,,\\
\end{split}
\end{equation}
where the term $\int_0^\infty dt \langle H^{(\alpha)}(t)\rangle^{eq}\langle H^{(\beta)}(0)\rangle^{eq}$ vanishes. After Fourier transforming in time and using the definition $\langle E_i (\rr,t)  E_j (\rr',0) \rangle^{eq}=\int \frac{d\omega}{2\pi} e^{-i\omega t} \langle E_i(\textbf{r})E_j^*(\textbf{r}')\rangle^{eq}_\omega $,
the first integrand in Eq.~\eqref{eq:Wick} reads (the other one is treated analogously),
\begin{equation}\label{FourierTransform}
\begin{split}
&\int_0^\infty dt \langle E_i (\rr,t)  E_j (\rr',0) \rangle^{eq}\langle J_i (\rr,t)  J_j (\rr',0) \rangle^{eq}=\\
&\int_0^\infty \frac{d\omega}{2\pi}\langle E_i (\rr)  E_j^* (\rr') \rangle_\omega^{eq}\langle J_i (\rr)  J^*_j (\rr')
\rangle_{-\omega}^{eq}\,.\\
\end{split}
\end{equation}
The equilibrium spectral density $\langle E_i(\textbf{r})E_j^*(\textbf{r}')\rangle^{eq}_\omega$ of the electric field is well-known and can be expressed via the dyadic retarded Green's function $G_{ij}$ of the system (a form of the fluctuation-dissipation theorem~\cite{Eckhardt,KuboBook,KuboFDT}),
\begin{equation}\label{FDT}
\begin{split}
\langle E_i (\rr) {E}^*_j(\rr')\rangle_\omega^{eq}=\frac{8\pi \hbar }{1-e^{-\hbar \omega/k_BT}}\frac{\omega^2}{c^2} \Im G_{ij}(\rr,\rr';\omega).
\end{split}
\end{equation}
This Green's function is straightforwardly found for a two component  system~\cite{LongPaper} (where we employ operator notation $\mathbb{G}\hat{=}G_{ij}({\bf r},{\bf r}')$) as
\begin{equation}\label{Green}
\G=\left(1+\GO\Td\right)\frac{1}{1-\GO\To\GO\Td}\left(1+\GO\To\right)\GO\,.
\end{equation}
Here, $\mathbb{T}_\alpha=\mathbb{T}_{\alpha, ij}({\bf r},{\bf r}')$ is the T-operator of $\alpha$, relating the scattered wave to an incoming wave of unit amplitude~\cite{Jamal}; $\GO$ is the Green's function of free space, i.e., the solution of the free space Helmholtz's equation, which relates
the total field and the total current, as used in Eq.~\eqref{CurrentFlux}, by
\begin{equation}
E_i(\omega)=4\pi i\frac{\omega}{c^2} G_{0,ij} J_j(\omega)\,.\label{current}
\end{equation}
For a  single object with operator $\mathbb{T}$, the Green's function reduces accordingly to $\mathbb{G}=(1+\GO\T)\GO$~\cite{Jamal}. 

After some computation steps, we find a closed form for the correlation function in Eq.~\eqref{RHS} in terms of $\GO$ and the T-operators
of the entities, see Eq.~\eqref{eq:1}. One important step is that the integrals in Eq.~\eqref{RHS} can eventually be taken over all
space (due to the fact that $\T_\alpha=\T_\alpha({\bf r},{\bf r}')$ is only nonzero if both arguments are within
$V_\alpha$~\cite{Jamal,LongPaper}), such that together with the summation over vector index $i$, an operator trace arises. A comparison to the previously computed radiative heat transfer $\langle H^{(\beta)}\rangle$~\cite{LongPaper} (see also Eq.~\eqref{B1}), denoting the energy absorbed by object $\beta$ in the {\it non-equilibrium} situation with $T_1$, $T_2$ and $T_{env}$ unequal, explicitly shows the following equality 
\begin{equation}\label{result1}
\begin{split}
k^{(\beta)}_\alpha&\equiv-\left.\frac{d\langle H^{(\beta)}\rangle}{dT_{\alpha}}\right|_{\{T_{\alpha}\}=T_{env}=T}\\&=\frac{1}{k_B T^2} \int_{0}^\infty dt\langle H^{(\alpha)} (t) H^{(\beta)}(0) \rangle^{eq}.
\end{split}
\end{equation}
Here we define the linear radiative heat transport coefficient $k^{(\beta)}_\alpha$, as a measure of the change in the heat
absorption $\langle  H^{(\beta)}\rangle$ by object $\beta$ in response to a small change in temperature of $\alpha$. It is
interesting to note that for $\alpha\not=\beta$, Eq.~\eqref{result1} implies a nonlocal correlation between fluctuations in the different
objects, in contrast to the purely local character of Eq.~\eqref{GK}.

As a side note, Eq.~\eqref{result1} directly shows the positivity of the linear transport coefficient $k_\alpha^{(\alpha)}$, as
equilibrium auto-correlation functions have non-negative Fourier transforms~\cite{McLennanBook}. On the other hand,
Eq.~\eqref{result1} for $\alpha\neq\beta$ does not allow us to make a statement about the sign of $-k_\alpha^{(\beta)}$, which however is  non-negative as well~\cite{LongPaper,RodriguezJohnson}.

\begin{figure}\centering
\centering
\includegraphics[width=8.1 cm]{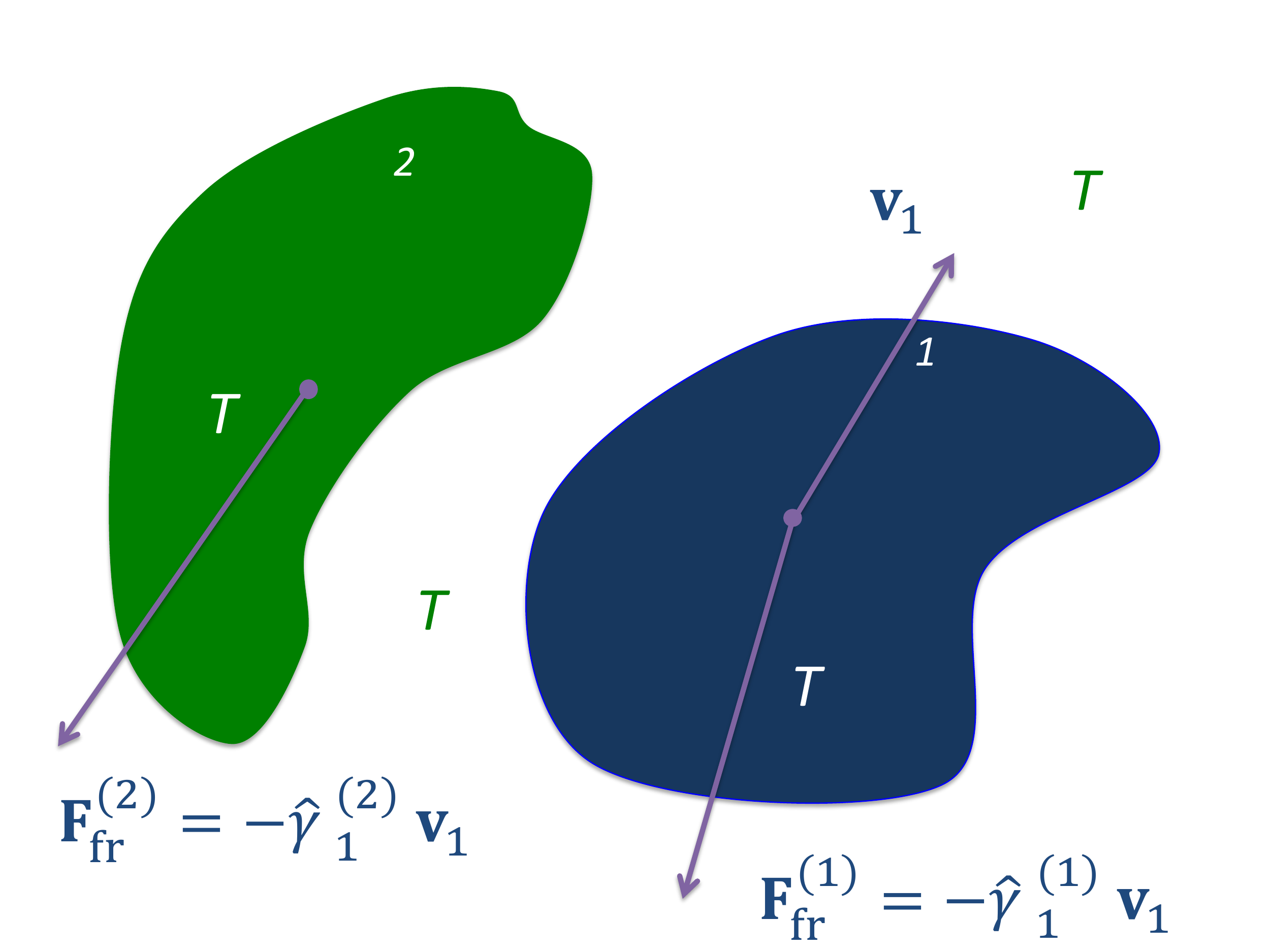}
\vspace*{-0.5 cm}
\caption{An object moving with velocity $\textbf{v}_{1}$ in the presence of a static object, gives rise to the vacuum friction
$-\hat{\gamma}_1^{(1)}\textbf{v}_1\equiv \frac{d\langle {\bf F}^{(1)} \rangle}{d {\textbf{v}}_{1}}|_{{\textbf{v}}_{1}=\bf{ 0}} \textbf{v}_1$ acting on
the moving object, and the force $-\hat{\gamma}_1^{(2)}\textbf{v}_1\equiv \frac{d\langle {\bf F}^{(2)} \rangle}{d {\textbf{v}}_{1}}|_{{\textbf{v}}_{1}=\bf{ 0}} \textbf{v}_1$  acting on the static one.}
\label{friction} 
\end{figure}
\vspace{0.5cm}
\subsection{Casimir force}\label{section3}
Now consider the change in the force $\textbf{F}^{(\beta)}$~\cite{Casimir,Antez,MK1,Jamal,LongPaper} when all objects are at rest, but with one temperature perturbed to non-equilibrium, as in Fig.~\ref{fig}.
We find that variations in force are related to the equilibrium correlation function of heat flux and force (compare to Eqs.~\eqref{eq:2} and \eqref{B4}) by
\begin{align}\label{result2}
&\left.\frac{d\langle\textbf{F}^{(\beta)}\rangle}{dT_\alpha}\right|_{\{T_\alpha\}=T_{env}=T}\notag\\&=-\frac{1}{k_B T^2} \int_{0}^\infty dt\langle \mathbf{F}^{(\beta)} (t) H^{(\alpha)}  (0) \rangle^{eq}\,.
\end{align}
This relation is found by steps analogous to the ones above Eq.~\eqref{result1}, starting from the Lorentz force acting on $\beta$, given by the volume integral
\begin{equation}\label{VolumeForce}
\begin{split}
F_i^{(\beta)}(t)=&\frac 1 c\int_{\rr\in V_\beta} d^3\rr  ~ \varepsilon_{ijk} \{J_j(\rr,t)\,, B_k(\rr,t)\}_S,\\
\end{split}
\end{equation}
where $B_k$ is the $k$-th component of the magnetic field, and $\varepsilon_{ijk}$ is the Levi-Civita symbol. As before, the equality in Eq.~\eqref{result2} is established by direct comparison to the result for the Casimir force in the non-equilibrium situation with $T_1$, $T_2$ and $T_{env}$ unequal given by Eq.~\eqref{B3}~\cite{LongPaper}. (See Eq.~\eqref{eq:2} for the explicit result of the correlation function in Eq.~\eqref{result2}.)

The relation \eqref{result2} is anticipated from linear response in the density matrix, yielding the time integral containing the energy dissipation~\cite{McLennan} (in our case $H$). The awaited general relation for observable $\mathcal{O}(t)$
\begin{equation}\label{McLen}
\left.\frac{d \langle  \mathcal{O} \rangle}{dT_\alpha}\right|_{T_\alpha=T}=-\frac{1}{k_B T^2} \int_{0}^\infty dt\langle  \mathcal{O}(t)  H^{(\alpha)} (0) \rangle^{eq}\,,
\end{equation}
is however yet unproven in this framework. 

\section{Perturbing Velocity}\label{section4}
\subsection{Casimir force (vacuum friction)}\label{section41}
The equilibrium system can also be perturbed by moving object(s) $\alpha$
with a small velocity ${\textbf{v}}_{\alpha}$. The corresponding change
in the Casimir force acting on $\beta$, expressed in terms of the linear force coefficient $\hat{\gamma}_\alpha^{(\beta)}\equiv -\frac{d\langle {\bf F}^{(\beta)} \rangle}{d {\textbf{v}}_{\alpha}}|_{{\textbf{v}}_{\alpha}=\bf{ 0}}$ (see Fig.~\ref{friction}), is
related to the auto-correlation function of the Casimir force in equilibrium~\cite{Reynaud, Reynaud93, Volokitin2}, in analogy to the Kirkwood formula in Eq.~\eqref{Kirkwood} (the diagonal part $\hat{\gamma}_\alpha^{(\alpha)}$ is the friction coefficient of $\alpha$). Here, we explicitly confirm this relation for the fluctuating electromagnetic field, thereby providing a closed expression for the vacuum thermal friction. We find, elaborating in analogy to the derivation of Eqs.~\eqref{result1} and~\eqref{result2}, for the fluctuations of the Casimir force,

\begin{widetext}
\begin{equation}\label{FF}
\begin{split}
&\left(\hat{\gamma}_\alpha^{(\beta)}\right)_{ij}=\frac{1}{k_B T}\int_{0}^\infty dt\langle \delta F_i^{(\beta)} (t) \delta F_j^{(\alpha)}(0) \rangle^{eq}=-\frac{\hbar^2}{\pi k_BT}\int_0^\infty d\omega\frac{ e^{\hbar \omega /k_BT}}{\left(e^{\hbar \omega /k_BT}-1\right)^2} \\&\Im\rm{Tr} \left\{ \partial_i (1+\mathbb{G}_0 \mathbb{T}_{\overline{\alpha}})  \frac{1}{1-\mathbb{G}_0 \mathbb{T}_\alpha\mathbb{G}_0 \mathbb{T}_{\overline{\alpha}}} \mathbb{G}_0 [\textit{i}(\partial_j \mathbb{T}_\alpha^*-\mathbb{T}_{{\alpha}}\partial_j)-2\mathbb{T}_\alpha\partial_j\rm{Im}[\mathbb{G}_0]\mathbb{T}_\alpha^*]\frac{1}{1-\mathbb{G}_0^* \mathbb{T}_{\overline{\alpha}}^*\mathbb{G}_0^* \mathbb{T}_\alpha^*}\left(\delta_{\alpha\beta}+\delta_{\overline{\alpha}\beta}\mathbb{G}_0^*\mathbb{T}_{\overline{\alpha}}^*\right)\right\},
\end{split}
\end{equation}
\end{widetext}
where ${\overline{\alpha}}=1$ if $\alpha=2$ and vice versa. Note that the matrix $\left(\hat{\gamma}_\alpha^{(\beta)}\right)_{ij}$ has in general non-zero off-diagonal elements in $ij$,
and the force need not be parallel to the velocity. 
While Eq.~\eqref{FF} contains both the thermal and zero point contributions
to the net Casimir force, at $T=0$K, the linear force coefficient $\hat{\gamma}_\alpha^{(\beta)}$ vanishes, and there is no linear response in velocity.

There is, however, response related to higher time derivatives of displacement, 
in accordance with known results (see, e.g. Refs.~\cite{Pendry, Reynaud, Volokitin2,Mohammad}). The friction coefficient in Eq.~\eqref{FF} has been previously computed for the special cases of two parallel plates, and 
for a small particle in front of a plate~\cite{Volokitin2}. 

The first equality sign in Eq.~\eqref{FF} can be confirmed by deriving the linear force coefficient directly.  Then the term $-i\mathbb{G}_0 [\textit{i}(\partial_j \mathbb{T}_\alpha^*-\mathbb{T}_\alpha\partial_j)-2\mathbb{T}_\alpha\partial_j\rm{Im}[\mathbb{G}_0]\mathbb{T}_\alpha^*]\mathbb{G}_0^*$ is found as the disturbed field correlator due to the moving object given by Eq.~\eqref{C8}~\cite{Henkel}. This field then undergoes scattering due to the surrounding objects, and computing the force introduces another gradient, $\partial_i$, in analogy to Eq.~\eqref{B4} and Ref.~\cite{LongPaper}.

In case of an isolated object, the friction tensor $\hat{\gamma}_\alpha^{(\alpha)}$ simplifies (where we omit the label $\alpha$) to
\begin{equation}\label{FFsingle}
\begin{split}
&\hat{\gamma}_{ij}=\frac{1}{k_B T}\int_{0}^\infty dt\langle \delta F_i (t) \delta F_j (0) \rangle^{eq}=\frac{2\hbar^2}{\pi k_B T}\int_0^\infty d\omega\\& \frac{e^{\hbar \omega /k_BT}}{\left(e^{\hbar \omega /k_BT}-1\right)^2}\Im\rm{Tr} \left\{\partial_i (1+\mathbb{G}_0 \mathbb{T})\partial_j \Im[ \mathbb{G}_0] \mathbb{T}^*\right\}\,.\\
\end{split}
\end{equation}
This is equivalent to the force acting on the object at rest in a photon gas moving in direction $j$~\cite{Vanik2} (see also Eqs.~\eqref{B4} and \eqref{C8}). The latter has the electric field correlator $\propto -i\partial_j \Im[\mathbb{G}_0]$.

The trace in Eqs.~\eqref{FF} and~\eqref{FFsingle} can be readily  evaluated in any basis, as exemplified in detail in Ref.~\cite{LongPaper}
for  heat transfer and the non-equilibrium Casimir force. For example, Eq.~\eqref{FFsingle} reads for a sphere,
\begin{equation}\label{FFsphere}
\begin{split}
&\hat{\gamma}_{ij}=-\delta_{ij}\frac{2\hbar^2}{3\pi k_B T}\int_0^\infty d\omega \frac{ e^{\hbar \omega /k_BT}}{\left(e^{\hbar \omega /k_BT}-1\right)^2}\frac{\omega^2}{c^2} \\&\sum_{P,l,m} \Re\left[\mathcal{T}^P_l+3 a(l,m)^2 \mathcal{T}^P_l \mathcal{T}^{\overline{P}*}_l+6 b(l,m)^2  \mathcal{T}^P_l \mathcal{T}^{P*}_{l+1})\right]\,,\\
\end{split}
\end{equation}
where $\mathcal{T}_l^P\equiv \mathcal{T}_l^P(\omega,R)$ is the T-matrix element for the scattering of a spherical wave with frequency $\omega$, wave numbers $l,m$ and polarization $P=\{N,M\}$ from a sphere with radius $R$ (see e.g. 
Ref.~\cite{LongPaper} for the expressions for $\mathcal{T}$). $\overline{P}=N$ if $P=M$ and vice versa. The coefficients in Eq.~\eqref{FFsphere} are
\begin{align}
&a(l,m)=\frac{m}{l(l+1)}\,,\\
&b(l,m)=\frac{1}{l+1}\sqrt{\frac{l(l+2)(l-m+1)(l+m+1)}{(2l+1)(2l+3)}}\,.
\end{align}
Expanding Eq.~\eqref{FFsphere} to lowest order in $R$ (noting that $\mathcal{T}_l^P\propto R^3$ to lowest order), and by relating the term linear in $\mathcal{T}_l^P$ to the polarizability, we recover the result obtained in Ref.~\cite{Vanik2}. Interestingly, if the sphere is a perfect mirror (in which case it does not emit heat radiation), Eq.~\eqref{FFsphere} gives a finite result. Specifically, in the limit of a small spherical mirror, we find
\begin{equation}
\hat{\gamma}_{ij}=\delta_{ij}\frac{896\pi^7}{135}\frac{\hbar R^6}{\lambda_T^8},
\end{equation}
where $\lambda_T=\hbar c/k_B T$ is the thermal wavelength, indicating that the friction coefficient is proportional to $T^8$.

\subsection{Radiative heat transfer} \label{section42}
An additional anticipated general linear response relation that is complimentary to Eq.~\eqref{McLen} reads,

\begin{equation}\label{McLen2}
\left.\frac{d \langle  \mathcal{O} \rangle}{d\textbf{v}_\alpha}\right|_{\textbf{v}_\alpha=\textbf{0}}=-\frac{1}{k_B T} \int_{0}^\infty dt\langle   \mathcal{O} (t) \delta \mathbf{F}^{(\alpha)}(0) \rangle^{eq}\,.
\end{equation}
Consider $\mathcal{O}=H^{(\alpha)}$, then by comparing Eq.~\eqref{McLen2} to Eq.~\eqref{result2} we can finally provide the Onsager theorem for fluctuational electrodynamics by writing,
\begin{equation}\label{maxwell}
\left.\frac{d\langle H^{(\alpha)}\rangle}{d\textbf{v}_\beta}\right|_{\textbf{v}_\beta=0}=-T \left.\frac{d\langle\textbf{F}^{(\beta)}\rangle}{dT_\alpha}\right|_{\{T_\alpha\}=T_{env}=T}.
\end{equation}
Here we used the symmetry $\int_{0}^\infty dt\langle  H^{(\alpha)}(t) \delta \mathbf{F}^{(\beta)} (0) \rangle^{eq}=-\int_{0}^\infty dt\langle  \delta \mathbf{F}^{(\beta)} (t) H^{(\alpha)}(0)  \rangle^{eq}$, as found explicitly by using the methods outlined above.

\section{Experimental relevance and summary}\label{section5}
Let us finally comment on experimental relevance of the above results. 
While the friction in Eq.~\eqref{FF} is in principle measurable in precision
force experiments~\cite{ExpFriction}, the fluctuations of $H$ in Eq.~\eqref{result1} are  harder to access.
We propose instead a method for indirect detection based on equilibrium 
fluctuations of internal energy ${\cal{E}}^{(\alpha)}(t)$ from Eq.~\eqref{result1}. 
Energy conservation requires (in the absence of other heat sources) that
\begin{align}
\frac{\partial}{\partial t} \delta{\cal{E}}^{(\alpha)} (t)=H^{(\alpha)}(t)\,,\label{eq:Con}
\end{align}
using which Eq.~\eqref{result1} can be recast as
\begin{equation}\label{result11}
\begin{split}
k^{(\beta)}_\alpha&=-\frac{1}{k_B T^2}\lim_{t\to 0} \frac{\partial}{\partial t}\langle \delta {\cal{E}}^{(\alpha)} (0) \delta {\cal{E}}^{(\beta)}(t) \rangle^{eq}\,.
\end{split}
\end{equation}
Relations of this type are sometimes referred to as macroscopic fluctuation--dissipation conditions. The spectrum of energy fluctuations  of $\alpha$ in
the environment of other objects can be related to $k_\alpha^{(\alpha)}$ and its heat capacity $C^{(\alpha)}$. 
Omitting the index $\alpha$ for brevity, the equal time correlations of energy
are obtained by standard statistical physics arguments as 
$\langle \delta {\cal{E}}(0)^2 \rangle^{eq}=C k_B T^2$. 
Hence, by integrating Eq.~\eqref{result11} we obtain,
\begin{equation}\label{dEdE}
\langle \delta {\cal{E}}(t) \delta {\cal{E}}(0) \rangle^{eq}=C k_B T^2\left[1-\frac{k t}{C}+\cdots\right]\approx C k_B T^2 e^{-t/\tau}.
\end{equation}
The dots imply higher powers in $t$, which we have assumed lead to 
an overall exponential decay, with $\tau=C/k$. 
Thus, if the object's heat coupling to the remainder of the system is 
{\it dominated} by vacuum heat transfer $H$, then its internal energy 
will {\it fluctuate with timescale $\tau$}. 
The equilibrium Casimir force is a function of temperature.
If its fluctuations $\delta{\bf F}^{(\alpha)}(t)$ can be assumed to depend on 
$\delta {\cal{E}}^{(\alpha)}(t)$, then they should also exhibit 
a signature of the timescale $\tau$. 
Without needing to specify the explicit dependence of ${\bf F}^{(\alpha)}(t)$ on
$\delta {\cal{E}}^{(\alpha)}(t)$, we can thus claim that a Fourier-analysis of 
${\bf F}^{(\alpha)}(t)$ should reveal $\tau$ (besides other characteristic timescales), and hence provide an equilibrium means of detecting the vacuum heat conductivity. 
In order to fulfill Eq.~\eqref{eq:Con}, any mechanical contact to the object 
(e.g. by a cantilever) should be thermally insulated.
Furthermore, the relative fluctuations of energy are enhanced for smaller
$C$ (per Eq.~\eqref{dEdE}) favoring smaller objects.
For example, a setup of a silicon sphere of radius 1$\mu$m in front of
a silicon plate at a separation of 100nm, leads to a timescale 
of $\tau\approx 50\mu s$, which is large enough for experimental detection. 

To conclude, we have demonstrated that for a collection of well separated 
objects, there is a Green-Kubo matrix relating radiative heat transfers to
long-range \mbox{(cross-)correlations} of the heat flux fluctuations. 
A similar expression relates the non-equilibrium component of the Casimir force 
to the correlations between force and heat transfer in equilibrium. 
The vacuum (frictional forces) from thermal photons due to motion
of an object (or any collection of objects) can be written in a compact 
form using scattering theory. Finally, we provide Onsager's theorem for $H$ and $\bf{F}$.
The results are based on fluctuational electrodynamics which assumes 
that each body is separately in thermal equilibrium; 
an assumption that could potentially be investigated in future work, and
is expected to break down in far from equilibrium situations.

\section{Acknowledgements}

We thank C. Maes, G. Chan, G. Bimonte, T. Emig, R. L. Jaffe,
M. F. Maghrebi, M. T. H. Reid and N. Graham for helpful discussions. This research was supported by the DFG grant No. KR 3844/2-1, NSF Grant No. DMR-12-
06323, DOE grant No. DE-FG02- 02ER45977, and DAAD grant No. GR6212-A1280210.

\appendix
\section{Equilibrium correlations}
We consider a system of two arbitrary objects (or two sets of distinct objects) in equilibrium with the environment at temperature $T$. The objects' scattering properties are described by their scattering (T) operators $\mathbb{T}_1$ and $\mathbb{T}_2$, respectively. Then, following the derivation outlined in the main article, we obtain the following results for the desired correlation functions,
\begin{align}
\int_0^\infty dt\langle &H^{(1,2)}(t)H^{(2)}(0)\rangle^{eq}=\nonumber\\&\frac{2\hbar^2}{\pi}\int_0^\infty d\omega\frac{\omega^2e^{\hbar\omega/k_BT}}{(e^{\hbar\omega/k_BT}-1)^2}\Im \rm{Tr} ~\mathbb{M}^{(2)}_{1,2}\, ,\label{eq:1}\\
\int_0^\infty dt \langle &H^{(1,2)}(t)\textbf{F}^{(2)}(0)\rangle^{eq}=\nonumber\\&-\frac{2\hbar^2}{\pi}\int_0^\infty d\omega\frac{\omega e^{\hbar\omega/k_BT}}{(e^{\hbar\omega/k_BT}-1)^2}\Re \rm{Tr}~ \nabla\mathbb{M}^{(2)}_{1,2}\, , \label{eq:2}
\end{align}
where we have introduced the operators
\begin{align}
&\mathbb{M}_1^{(2)}=(1+\mathbb{G}_0 \mathbb{T}_2)  \frac{1}{1-\mathbb{G}_0 \mathbb{T}_1\mathbb{G}_0 \mathbb{T}_2} \nonumber\\&\times\mathbb{G}_0 [\Im[\mathbb{T}_1]-\mathbb{T}_1\rm{Im}[\mathbb{G}_0]\mathbb{T}_1^*]\mathbb{G}_0^*\frac{1}{1- \mathbb{T}_2^*\mathbb{G}_0^* \mathbb{T}_1^*\mathbb{G}_0^*}\mathbb{T}_2^*\, ,, \label{eq:3}\\
&
\mathbb{M}_2^{(2)}=(1+\mathbb{G}_0 \mathbb{T}_1)  \frac{1}{1-\mathbb{G}_0 \mathbb{T}_2\mathbb{G}_0 \mathbb{T}_1}\nonumber\\& \times\mathbb{G}_0 [\Im[\mathbb{T}_2]-\mathbb{T}_2\rm{Im}[\mathbb{G}_0]\mathbb{T}_2^*]\frac{1}{1-\mathbb{G}_0^* \mathbb{T}_1^*\mathbb{G}_0^* \mathbb{T}_2^*}\,. \label{eq:4}
\end{align}

\section{Heat transfer and Casimir force}
We summarize the relevant results for heat transfer and non-equilibrium Casimir forces from Ref.~\cite{LongPaper}. Consider objects $1$ and $2$  held at   temperatures $T_1$ and $T_2$, and with the environment  at temperature $T_{env}$. The heat absorbed by object $2$ is given by Eq.~(69) from Ref.~\cite{LongPaper} as
\begin{equation}\label{B1}
\langle H^{(2)}\rangle(T_1,T_2,T_{env})=\sum_{\alpha=1,2}\langle H_\alpha^{(2)}\rangle({T_\alpha}) -\langle H_\alpha^{(2)}\rangle(T_{env})\,.
\end{equation}
Here, $\langle H_1^{(2)}\rangle$ is the heat transfer from object $1$ to object $2$, 
and $\langle H_2^{(2)}\rangle $ is the so-called self-emission by object $2$, corresponding to the heat lost by object $2$ due to the presence of object $1$. These are given by Eqs.~(56) and~(65) from Ref.~\cite{LongPaper} respectively,
\begin{align}\label{B2}
&\langle H_{1,2}^{(2)}\rangle(T_{1,2})=-\frac{2\hbar}{\pi}\int_0^\infty \frac{\omega d\omega}{e^{\hbar \omega /k_B T_{1,2}}-1} \Im\rm{Tr}~\mathbb{M}_{1,2}^{(2)}\,.
\end{align}
The Casimir force acting on an arbitrary object $2$ is given by Eq.~(79) in Ref.~\cite{LongPaper}, and can be written as
\begin{equation}\label{B3}
\begin{split}
\langle \textbf{F}^{(2)}\rangle({T_1,T_2,T_{env}})=&\langle \textbf{F}^{(2)}\rangle ^{eq}(T_{env})\\&+\sum_{\alpha=1,2}[\langle \textbf{F}_\alpha^{(2)}\rangle({T_\alpha}) -\langle \textbf{F}_\alpha^{(2)}\rangle({T_{env}})]\,.
\end{split}
\end{equation}
The equilibrium Casimir force $\langle \textbf{F}^{(2)}\rangle ^{eq}$ is much studied~\cite{Jamal}, and not relevant for our analysis. The non-equilibrium contribution $\langle \textbf{F}_{1}^{(2)}\rangle$ acts on object $2$ due to the sources in object $1$. The other non-equilibrium contribution is the self-force, $\langle \textbf{F}_{2}^{(2)}\rangle $, and represents the force that acts on object $2$ due to the sources in the  object itself. These non-equilibrium contributions to the Casimir force are given by Eqs.~(76) and~(77) in Ref.~\cite{LongPaper} as
\begin{align}\label{B4}
&\langle \textbf{F}_{1,2}^{(2)}\rangle(T_{1,2})=\frac{2\hbar}{\pi}\int_0^\infty \frac{ d\omega}{e^{\hbar \omega /k_BT_{1,2}}-1} \Re\rm{Tr}~ \nabla \mathbb{M}_{1,2}^{(2)}\,.
\end{align}
With Eqs.~\eqref{eq:1},\eqref{eq:2},\eqref{B1}-\eqref{B4}, the relations \eqref{result1} and \eqref{result2} of the main text can be confirmed.

\section{Field correlations sourced by a moving object}
Here we compute the spectral density $\mathbb{C}^{obj}(\textbf{r},\textbf{r}')\equiv\langle E_i(\textbf{r})E_j^*(\textbf{r}')\rangle^{obj}_{\omega}$ resulting from an isolated object moving with velocity $\textbf{v}$, to linear order in velocity. 
(In contrast to Eq.~(7) in the main text,  the correlator $\langle E_i(\textbf{r},t)E_j^*(\textbf{r}',0)\rangle$ is symmetrized.) 
Without loss of generality, consider the object moving along the $p$-axis, 
so that $\textbf{v}=v\hat{e}_p$.

We first consider an arbitrary equilibrium situation viewed in a reference frame moving with velocity $\textbf{v}$, which follows from the covariant treatment in 
Ref.~\cite{Henkel}. The spectral density  can be expressed in terms of the system's Green's function as
\begin{widetext}
\begin{equation}\label{C1}
C_{ij}(\textbf{r},\textbf{r}')\equiv\langle E_i(\textbf{r})E_j^*(\textbf{r}')\rangle_\omega=\int \frac{d^3 \textbf{k}}{(2\pi)^3} \int \frac{d^3 \textbf{h}}{(2\pi)^3} e^{i (\textbf{k} \cdot \textbf{r}+\textbf{h}\cdot \textbf{r}')} C_{ij}(\textbf{k},\textbf{h})\equiv\int \frac{d^3 \textbf{k}}{(2\pi)^3} \int \frac{d^3 \textbf{h}}{(2\pi)^3} e^{i (\textbf{k} \cdot \textbf{r}+\textbf{h}\cdot \textbf{r}')} \langle E_i(\textbf{k})E_j^*(\textbf{h})\rangle_{\omega},
\end{equation}
\end{widetext}
where we use the symmetric version of expression for $C_{ij}(\textbf{k},\textbf{h})$ from Ref.~\cite{Henkel},
\begin{widetext}
\begin{equation}\label{low}
\begin{split}
C_{ij}(\textbf{k},\textbf{h})=-\mathrm{sgn}(\omega)\frac{2i\pi\hbar\omega^2}{c^2}\left\{\coth\left(\frac{\hbar(\omega-k_p v)}{2k_BT}\right) G_{ij}(\omega,\textbf{k},\textbf{h})-{ \coth\left(\frac{\hbar(\omega+h_p v)}{2k_BT}\right)  G_{ji}^*(\omega,-\textbf{h},-\textbf{k})}\right\}\,.
\end{split}
\end{equation}
\end{widetext}
We have set $k_p=\textbf{k}\cdot\hat{e}_p$, and $G_{ij}(\omega,\textbf{k},\textbf{h})$ is the spatial/temporal Fourier transform of the Green's function $G_{ij}(t,\textbf{r},\textbf{r}')$ for the system. Note that for $v=0$ the equilibrium correlator in the rest frame is recovered. By expanding the field correlations to linear order in $v$, we obtain
\begin{equation}\label{C3}
\begin{split}
\left.\frac{dC_{ij}(\textbf{k},\textbf{h})}{dv}\right|_{v=0}=&-\mathrm{sgn}(\omega)\frac{4i\pi\hbar^2\omega^2}{c^2k_B T}
\frac{e^{\hbar\omega/k_BT}}{(e^{\hbar\omega/k_BT}-1)^2}
 \\&\times\left(k_p G_{ij}(\omega,\textbf{k},\textbf{h})+ {h_p}G_{ji}^*(\omega,-\textbf{h},-\textbf{k})\right)\,.
\end{split}
\end{equation}
Transforming back to real space we get for the Lorentz transformed field correlator to linear in $v$,
\begin{equation}\label{main1}
\begin{split}
\left.\frac{dC_{ij}(\textbf{r},\textbf{r}')}{dv}\right|_{v=0}=&-\mathrm{sgn}(\omega)\frac{4\pi\hbar^2\omega^2}{c^2k_B T}
\frac{e^{\hbar\omega/k_BT}}{(e^{\hbar\omega/k_BT}-1)^2}
\\&\times \left(\partial_{p} G_{ij}(\omega,\textbf{r},\textbf{r}')+\partial_{p'} G_{ji}^*(\omega,\textbf{r}',\textbf{r})\right)\,.
\end{split}
\end{equation}
For an isolated object,  Eq.~\eqref{main1} enables computing the field correlator $\mathbb{C}$ in a frame which is moving with respect to both the object and the environment. It follows with the Green's function of the system, expressed in terms of the object's T-operator, $\mathbb{G}=\mathbb{G}_0+\mathbb{G}_0\mathbb{T}\mathbb{G}_0$, and reads
\begin{equation}\label{eq:eqmov}
\begin{split}
\left.\frac{d\mathbb{C}}{dv}\right|_{v=0}=&-\mathrm{sgn}(\omega)\frac{4i\pi\hbar^2\omega^2}{c^2k_B T}
\frac{e^{\hbar\omega/k_BT}}{(e^{\hbar\omega/k_BT}-1)^2}\\&\times
 \left[2\partial_p\Im[\mathbb{G}_0]-i(\mathbb{G}_0\partial_p\mathbb{T}\mathbb{G}_0-\mathbb{G}_0^*\mathbb{T}^*\partial_p\mathbb{G}_0^*)\right]\,.
\end{split}
\end{equation}
To linear order in $v$, the result in Eq.~\eqref{eq:eqmov} can alternatively be found by a decomposition into two terms: one arising from the motion of the empty environment (with the static object present) and the other one resulting from the moving object in a static environment,
\begin{equation}\label{general}
\left.\frac{d\mathbb{C}}{dv}\right|_{v=0}=\left.\frac{d(	\mathbb{C}^{env}+\mathbb{C}^{obj})}{dv}\right|_{v=0}\,.
\end{equation}
We are interested in the latter component, $\mathbb{C}^{obj}$. The field sourced by the moving environment in the presence of a static object is computed by first considering Eq.~\eqref{main1} for the empty environment (described by $\mathbb{G}_0$), and then scattering at the static object~\cite{LongPaper}, to get
\begin{equation}\label{eq:l}
\begin{split}
\left.\frac{d\mathbb{C}^{env}}{dv}\right|_{v=0} =&-\mathrm{sgn}(\omega)\frac{8i\pi\hbar^2\omega^2}{c^2k_B T}
\frac{e^{\hbar\omega/k_BT}}{(e^{\hbar\omega/k_BT}-1)^2}\\&\times
 (1+\mathbb{G}_0\mathbb{T})\partial_p\Im[\mathbb{G}_0](\mathbb{T}^*\mathbb{G}_0^*+1)\,.
\end{split}
\end{equation}
The desired correlator can now be found by use of Eqs.~\eqref{eq:eqmov}, \eqref{general} and \eqref{eq:l}, and reads
\begin{equation}\label{C8}
\begin{split}
\left.\frac{d\mathbb{C}^{obj}}{dv}\right|_{v=0}=&-\mathrm{sgn}(\omega)\frac{4i\pi\hbar^2\omega^2}{c^2 k_B T}
\frac{e^{\hbar\omega/k_BT}}{(e^{\hbar\omega/k_BT}-1)^2}\\&\times\mathbb{G}_0\left[i(\partial_p \mathbb{T}^*-\mathbb{T}\partial_p)- 2\mathbb{T}\partial_p\Im[\mathbb{G}_0]\mathbb{T}^*\right]\mathbb{G}_0^*\,.
\end{split}
\end{equation}
This is precisely the source term in the expression of Eq.~(14) in the main text, demonstrating the equivalence of the force correlator in Eq.~(14) and the hereby found linear response result.

\bibliographystyle{apsrev4}

\end{document}